\documentclass[prb, groupedaddress, twocolumn, showpacs, floatfix]{revtex4-1}
	\usepackage{graphics}
	\usepackage{graphicx}
	\usepackage{epstopdf}
	\usepackage[caption=false]{subfig}
	\usepackage{textcomp}
	\usepackage{amssymb}
	\usepackage{amsmath}
       \usepackage{verbatim}
	\usepackage{color}
	\usepackage{hyperref}
	\usepackage{array}
	\newcolumntype{L}[1]{>{\raggedright\let\newline\\\arraybackslash\hspace{0pt}}m{#1}}
	\newcolumntype{C}[1]{>{\centering\let\newline\\\arraybackslash\hspace{0pt}}m{#1}}
	\newcolumntype{R}[1]{>{\raggedleft\let\newline\\\arraybackslash\hspace{0pt}}m{#1}}

\begin{document}

\title{Parity effect and single-electron injection for Josephson-junction chains deep in the insulating state}

\author{K.~Cedergren$^{1}$, S.~Kafanov$^{1}$, J-L.~Smirr$^{1}$, J.~H.~Cole$^{2}$ and T.~Duty$^{1}$}

	\affiliation{$^1$Centre for Engineered Quantum Systems (EQuS), School of Physics, University of New South Wales, Sydney 2052, Australia}
	\affiliation{$^2$ Chemical and Quantum Physics, School of Applied Sciences, RMIT University, Melbourne 3001, Australia}


\begin{abstract}

We have made a systematic investigation of charge transport in 1D chains of Josephson junctions where the characteristic Josephson energy is much less than the single-island Cooper-pair charging energy, $E_\mathrm{J}\ll E_{CP}$.  Such chains are deep in the insulating state, where superconducting phase coherence across the chain is absent, and a voltage threshold for conduction is observed at the lowest temperatures. We find that Cooper-pair tunneling in such chains is completely suppressed. Instead, charge transport is dominated by tunneling of single electrons, which is very sensitive to the presence of BCS quasiparticles on the superconducting islands of the chain.  Consequently we observe a strong parity effect, where the threshold voltage vanishes sharply at a characteristic parity temperature $T^*$, which is significantly lower than the the critical temperature, $T_c$. A measurable and thermally-activated zero-bias conductance appears above $T^*$, with an activation energy equal to the superconducting gap, confirming the role of thermally-excited quasiparticles. Conduction below $T^*$ and above the voltage threshold occurs \textit{via} injection of single electrons/holes into the Cooper-pair insulator, forming a non-equilibrium steady state with a significantly enhanced effective temperature. Our results explicitly show that single-electron transport dominates deep in the insulating state of Josephson-junction arrays. This conduction process has mostly been ignored in previous studies of both superconducting junction arrays and granular superconducting films below the superconductor-insulator quantum phase transition. 

\end{abstract}
\maketitle

\section{Introduction}

For several decades now, chains and arrays of low-capacitance Josephson junctions have attracted much attention as systems in which quantum phase transitions can be studied\cite{PhysRevB.30.1138.Bradley,PhysRep.355.235.Fazio}, and as many-body platforms that could enable novel quantum phases\cite{PhysRevLett.79.3736.Glazman, PhysRevLett.88.227005.Doucot} and topologically protected states\cite{PhysRevLett.90.107003.Doucot, Nature.415.503.Ioffe}. However, there have been many more theoretical proposals along these lines than experimental works, and experiments have mostly been confined to several intriguing avenues of research concerning the superconductor-insulator transition\cite{IntJModPhysB.24.4081.Goldman}, the dynamics of quantum-phase slips\cite{NaturePhys.6.589.Pop, PhysRevB.85.024521.Manucharyan, NewJPhys.15.095014.Ergul}, metrological current standards\cite{Nature.434.361.Bylander}, conjectured solitonic phenomena\cite{PhysRevB.54.6857R.Haviland, JLTP.124.291.Agren, PhysRevB.83.064517}, and the use of Josephson-junction chains as superinductors.\cite{Masluk_etal_2012} 

The insulating state of superconducting junction arrays is located below a superconductor-insulator (SI) quantum phase transition, and is synonymous with the destruction of superconducting phase coherence across the array, and localization of Cooper pairs. The Cooper-pair insulator occurs when the characteristic Cooper-pair charging energy significantly exceeds the Josephson coupling energy, $E_{CP}\gg E_\mathrm{J}$. The understanding and engineering of charge transport deep in the insulating state presents difficult problems due to the competition between various modes of transport which include both Cooper-pair and single-electron tunneling processes\cite{PhysRevB.76.020506R.Bylander, NewJPhys.16.063019.Cole}, both of which occur in the presence of significant disorder.

Nearly all studies of charge transport in insulating arrays, however, start from a point of view where it is assumed that the low energy excitations that play a role in conduction are Cooper pairs. This is based on the assumption that the temperature is sufficiently low to ignore the presence of unpaired electrons.  A recent calculation of the dc conductivity for arrays deep in the insulating state, based upon single Cooper-pair excitations and including weak disorder, was put forth by Syzranov \textit{et al}.\cite{PhysRevLett.103.127001.Syzranov}. Their model proposes Cooper pairs as the sole charge carriers, and describes transport in terms of variable range hopping between adjacent islands as a result of Josephson tunneling. In another theoretical study by Fistul \textit{et al.}\cite{Fistul_etal_2008}, a model for Cooper-pair transport in the insulating state of 1 and 2 D arrays of Josephson junctions is applied to the interpretation of experimental data from granular superconducting films. Similar lines of thinking have been taken in the analysis of Josephson-junction array experiments, also based on the assumption that charge transport far below the superconducting transition temperature, $T_c$, is predominantly carried by Cooper pairs \cite{PhysRevB.88.144506.Zimmer, ChowPRL98}. It is important to realize that in most experiments, the contribution of quasiparticles cannot be ignored. This contribution has proven important in studies of qubits \cite{CWang2014} and Cooper-pair transistors\cite{Aumentado2004}, but has only recently been studied theoretically in JJ chains \cite{arXiv:1503.01905}

We have made a systematic investigation of charge transport for temperatures ranging from $\mathrm{10\,mK}$ to  $\mathrm{1\,K}$ in 1D Josephson arrays that are characterized by large charging energies and high junction resistances. In this regime the arrays are deep in the insulating state. We find that Cooper-pair transport is completely suppressed and charge transport proceeds via single-electron injection into the Cooper-pair insulator. Furthermore, we observe a strong parity effect\cite{PhysRevLett.69.1993.Averin}, with a well-defined cross-over temperature $T^{*}$, at which the voltage injection threshold decreases sharply. The parity effect has been studied previously in superconducting single-electron transistors,\cite{PhysRevLett.69.1997.Tuominen} and in Cooper-pair boxes,\cite{LafargePRL93,LafargeNature93} where it appears as a temperature crossover from 2e to 1e periodicity in normalized gate charge. In contrast, our array measurements show the parity effect has a global effect on charge transport through the  whole array. The presence of $\sim$\,1 thermally-excited BCS quasiparticle per island in the array significantly enhances the tunneling rates of single electrons through the chain, and simultaneously destroys the insulating state of the array, as the voltage threshold for single-electron injection is suppressed to zero.

\section{Devices and Measurements}

We have fabricated 1D chains of Al-AlO$_x$-Al Josephson junctions having a length of $N$\,=\,50 junctions using electron beam lithography, followed by thermal shadow-evaporation and \textit{in-situ} oxidation of $\mathrm{Al}$ films, which have a thickness of 30\,nm. We have focused on three arrays with slightly different properties, having in common large junction resistances, $R_J \gg R_Q \equiv (2e)^2 / h$ and large Cooper-pair charging energies, $E_{CP}\equiv (2e)^2/2C_J \gg E_\mathrm{J}$, where $C_J$ is the junction capacitance, as described in Table 1. The arrays exhibit a Coulomb blockade at low temperatures, both in the superconducting state as well as in the normal state, which is obtained by suppressing the superconducting gap in the films using an external parallel magnetic field. 


\begin{table}
\begin{ruledtabular}
\caption{Device parameters}
\begin{tabular}{ C{0.13\linewidth}  C{0.16\linewidth}  C{0.17\linewidth}   C{0.13\linewidth}  C{0.14\linewidth}   C{0.15\linewidth}  }

device	& junction area ($\mu m)^2$ & island volume ($\mu m)^3$ & $R_J$ & $E_{CP}$	&  $E_\mathrm{J}$\\  
	
	\hline  A	 &  $0.015$	 &	$0.0029$  & 248  k$\Omega$ & 150\,$\mu$eV	  & 2.7\,$\mu$eV \\ 
	\hline  B 	&  $0.015$	&	$0.0029$ & 312  k$\Omega$	&  170\,$\mu$eV &  2.2\,$\mu$eV \\ 
	\hline  C	&  $0.003$	&      $0.0016$ &  786 k$\Omega$  &  680\,$\mu$eV &   0.86\,$\mu$eV \\

\end{tabular}
\end{ruledtabular}
\end{table}

The samples were bounded to a circuit board, mounted in a microwave tight Cu sample enclosure, and secured to the mixing chamber of a BlueFors LD400 cryogen-free dilution refrigerator with a base temperature of 10\,mK. Each DC line was filtered from high frequency radiation using three meters of ThermoCoax\textsuperscript{\textregistered}, thermally anchored at each stage of the dilution refrigerator, and having a measured low-pass cutoff frequency of $\sim$\,1 MHz. The lines were additionally filtered using chip LC components on the circuit board, and at room temperature using low-pass LC filters. Several measurements of superconducting single-electron transistors were made using this setup that clearly showed $2e$-periodic stability diagrams, indicating negligible quasi-particle poisoning due to the presence of non-equilibrium quasiparticles in the measurement leads. 

We have characterized the devices using current-voltage measurements with voltage biases ranging from 5\,$\mu$V to over 50\,$m$V, and with a DC current resolution as low as $\sim$\,0.8 femtoampere for currents up to several nanoamperes. In addition, we have used a parallel magnetic field $B_{||}$ to continuously suppress the superconducting gap $\Delta(B_{||})$. From our measurements we have found that the gap depends on the parallel magnetic field through $\Delta(B_{||})\simeq \Delta(0)(1-B_{||}^2/B_{c||}^2)$, as expected\cite{TinkhamBook}, with $\Delta(0)=210\pm 10\mu e\mathrm{V}$, and $B_{c||}=0.59\pm 0.02\,\mathrm{T}$.

It is important to distinguish between large-scale current-voltage characteristics (or $IVC$s) and small-scale $IVC$s. In the presence of a superconducting gap, `large-scale' refers to voltage biases that span the onset of direct quasiparticle tunneling due to pair breaking across every junction in the chain. This occurs for $eV \simeq 2N\Delta$. `Small-scale' bias in the superconducting state refers to voltages $V$  that are a substantially small fraction of  $2 N \Delta/e$, in which case conduction is also referred to as sub-gap transport. As the superconducting gap is suppressed below approximately twice the characteristic single-electron charging energy, $E_C=e^2/2C_J=E_{CP}/4$, the transition from small-scale to large scale occurs at $V \simeq E_C N/2e$, above which the Coulomb blockade is lifted across each junction. Our main results are primarily concerned with sub-gap transport; however, the large-scale $IVC$ measurements are used to experimentally characterize the energy scales and intrinsic disorder of the junction chains. 

\begin{figure}
\includegraphics[width=3.2in]{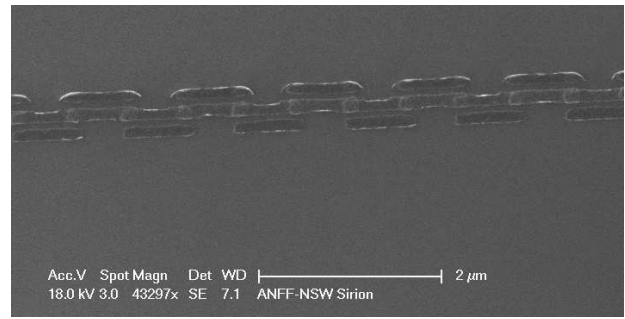}
\caption{SEM image of the center region of device A. The center row of islands forms a Josephson junction chain with well-defined tunnel junctions. The angle-evaporation process also produces shadow islands, which are not galvanically connected to the junction chain.
} 
\label{fig:SEM}
\end{figure}

\begin{figure}
   \subfloat
    {
        \includegraphics[width=3.3in]{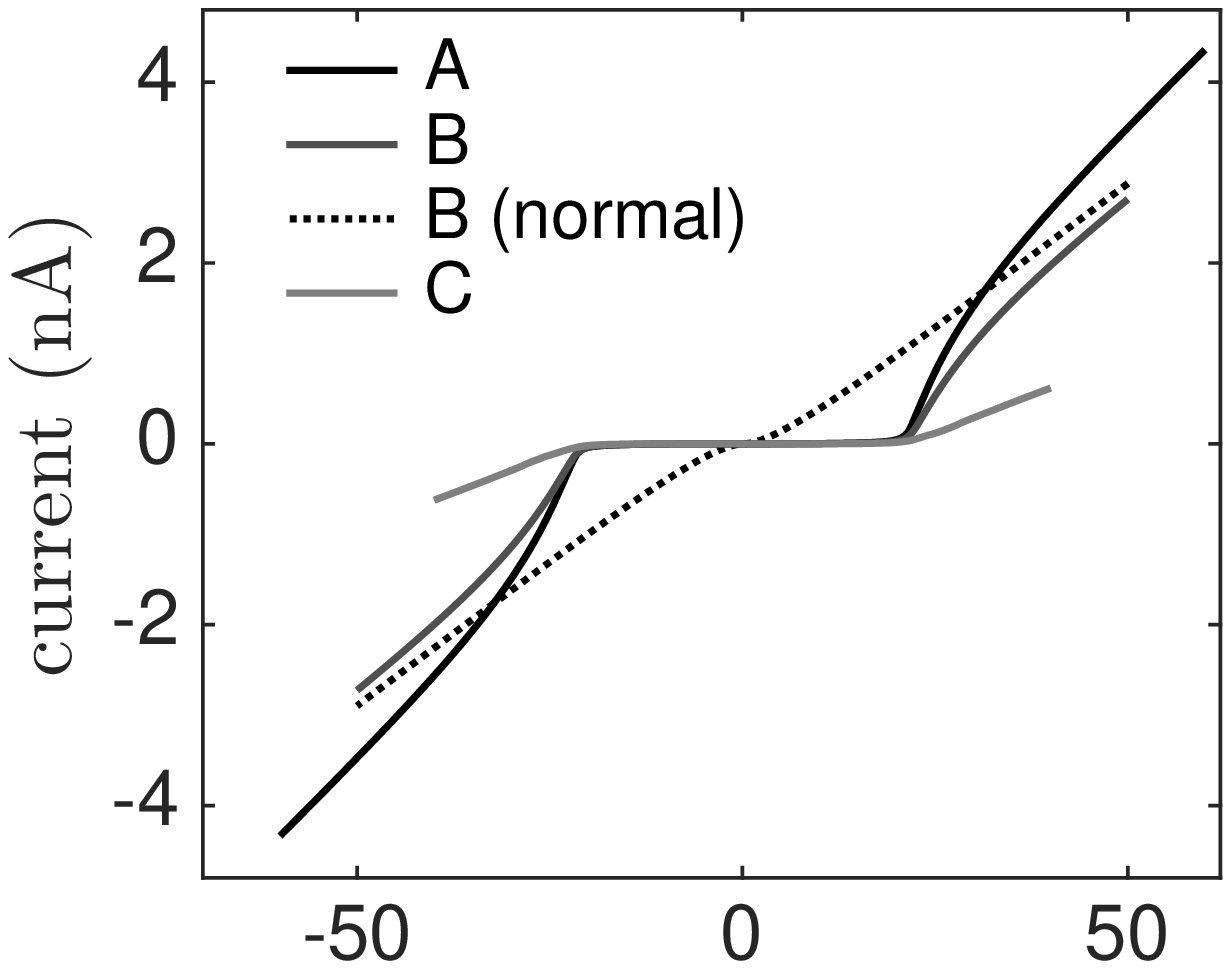}
        \label{fig:first_sub}
	\begin{picture}(0,0)
	\put(-115,+33){\includegraphics[height=2.3cm]{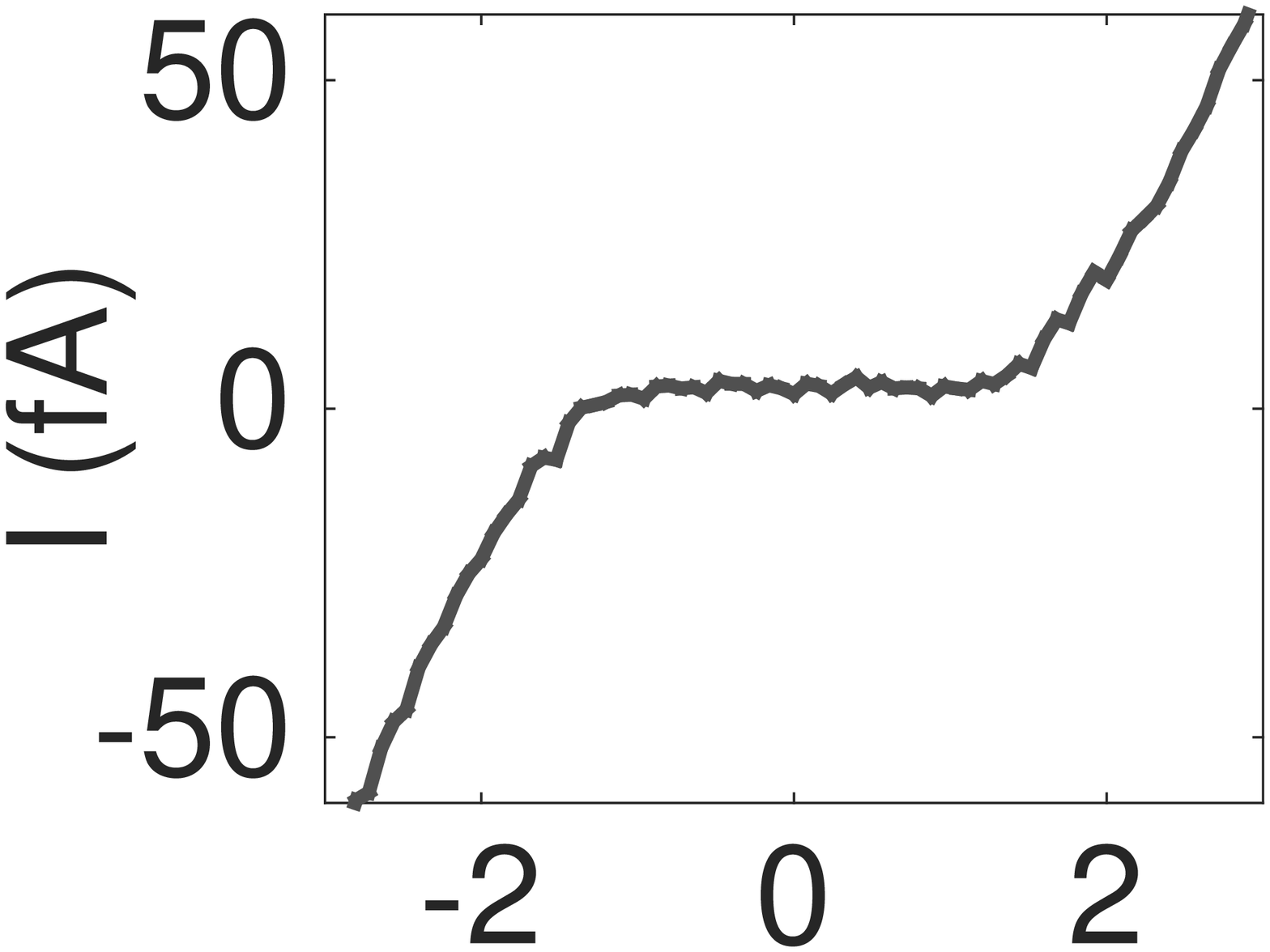}}
	\end{picture}

    }
    \\
\vspace{-18pt}
    \subfloat
    {
        \includegraphics[width=3.3in]{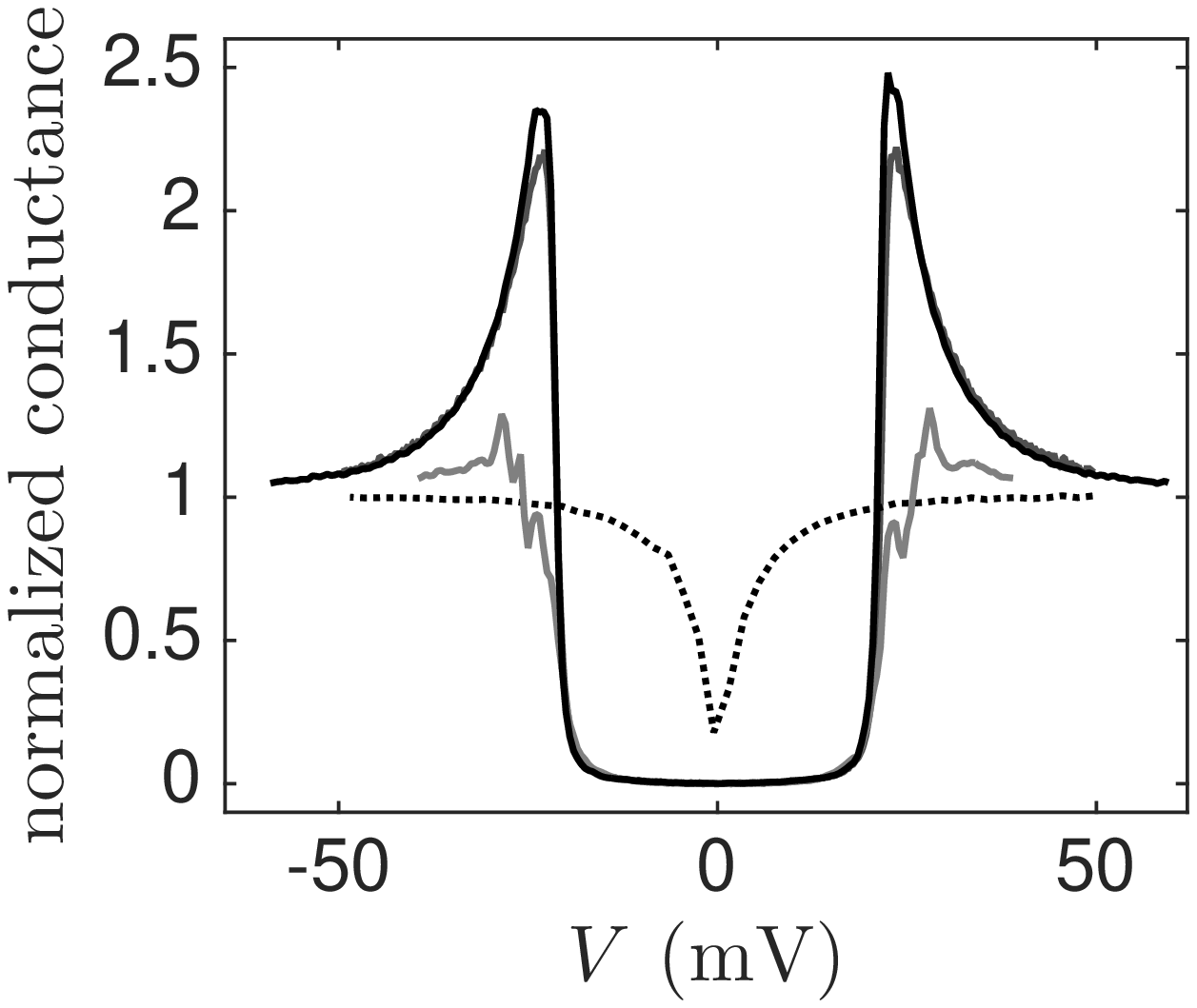}
        \label{fig:second_sub}
    }
 
 \caption{Large-scale current (upper plot) and conductance (lower plot) measurements for devices A, B and C in in the superconducting state for magnetic field $B_{||}$=0 (solid lines), and for device B in the normal state obtained with $B_{||}$=0.7 T (dotted lines). The inset in the upper plot shows the small-scale IVC at $T$\,=\,25\,mK for device B at femtoampere current resolution.  Conductance $(dI/dV)$ measurements in the lower plot are normalized to $(NR_J)^{-1}$ for each device.} 
    \label{fig:large_scale_small_scale_didv_vs_t_v7.fig}
\end{figure}

$IVC$ and differential conductance $(dI/dV)$ data are shown for the three devices of Table 1 in Figure 2. The asymptotic normal state conductance at large bias voltage $V$ determines the chain average of the normal state tunnel resistance $R_J$ per junction. From this, one can extract the average Josephson coupling energy using the Abegaokar-Baratoff relation $E_\mathrm{J}=\frac{1}{2}\Delta(R_Q/R_J)$. The Cooper-pair charging energy $E_{CP}\equiv (2e)^2/2C_J$ for each device is found experimentally by extrapolating its normal state $IVC$ data from the linear regime at large voltage bias to find the zero-current intercept $V_\mathrm{offset}=Ne/C_J$. $IVC$'s in the superconducting state in zero magnetic field show a steep onset of direct quasiparticle tunneling occurring at $V \simeq 2N\Delta_0/e$, allowing one to directly estimate $\Delta_0$. From the measurements in Figure 2, all devices are found to have nearly the same superconducting gap, $\Delta_0=210\pm10 \mu$eV.

Devices A and B show a large `BCS peak' that develops at $V \simeq 2N\Delta/e$. This peak arises from the overlap of divergent BCS quasiparticle density of states (DOS) of the superconducting islands of the chain. The peak is broadened primarily by offset charges that randomly shift the chemical potential of individual islands relative to their nearest neighbors. The result is a misalignment of the divergent BCS island DOS's across the chain, which in the absence of charge disorder would be aligned when the voltage bias across individual junctions equals $2\Delta/e$. The broadening and concomitant reduction of this BCS peak therefore gives a measure of the relative disorder of the chains due to offset charges and fabrication inhomogeneity.

As clearly seen in Figure 2, the BCS peak for device C is significantly more broadened as compared to devices A and B. In addition, there appears to be a random structure in the $IVC$ data for this device around the onset for direct quasiparticle tunneling. This is expected as device C has a significantly larger $E_{CP}$ compared to device A and B, which makes it much more sensitive to random offset charges. In addition, device C was fabricated in a separate processing run, using different lithographic development parameters, and therefore may have more intrinsic disorder due to reduced film and junction quality.

Small-scale $IVC$ data for device B is shown in the inset of Figure 2. Note that the current scale in the inset of Figure 2 is 5 orders of magnitude lower than in the main plot. There is a clear voltage threshold $V_t\simeq1.3\,\mathrm{mV}$ for the onset of femtoampere currents. Threshold voltages can be distinquished quite clearly in the differential conductance at lower temperatures, as shown in Figure  3. The region $|V| <  |V_t|$ shows a current blockade and a zero-bias conductance $G_0\equiv(dI/dV)_{V=0}$ that is identically zero, or at most, lower than our measurement resolution, $G^\mathrm{min}_0 = 10^{-12}\, \Omega^{-1}$. We experimentaly determine the voltage threshold $V_t$ as the absolute value of the voltage bias at which the conductance rises above $10^{-11}\,\Omega^{-1}$ (see lower plot in Figure 3), which is a factor of 10 greater than the measurement resolution. As seen in Figure 4, there is a characteristic temperature $T^*$ at which $V_t$ drops sharply to zero, and above which a measurable zero-bias conductance is observed.

A subset of the temperature dependent $IVC$ data for device B is shown in the lower plot of Figure 3 for $B_{||}=0$ and temperatures 25, 300, 350, 400 and 450\,mK. $T^*$ for device B in zero field is found to be 270\,mK (see Figure 4). The data at 300\,mK and above show a conductance peak around $V=0$ that grows with temperature, but starts out with a small mini-gap that appears to be a remnant of the blockade region below $T^*$. The conductance peak arises from the overlap of the BCS DOS from island to island across the chain, and only becomes evident when there are thermally excited quasiparticles occupying these states. (For a large Josephson junction such thermally excited quasiparticles give rise to a logarithmic singularity at $V=0$ in the $IVC$ at finite T).\cite{TinkhamBook}

\begin{figure}
  \subfloat
   {
    \includegraphics[width=3.3in]{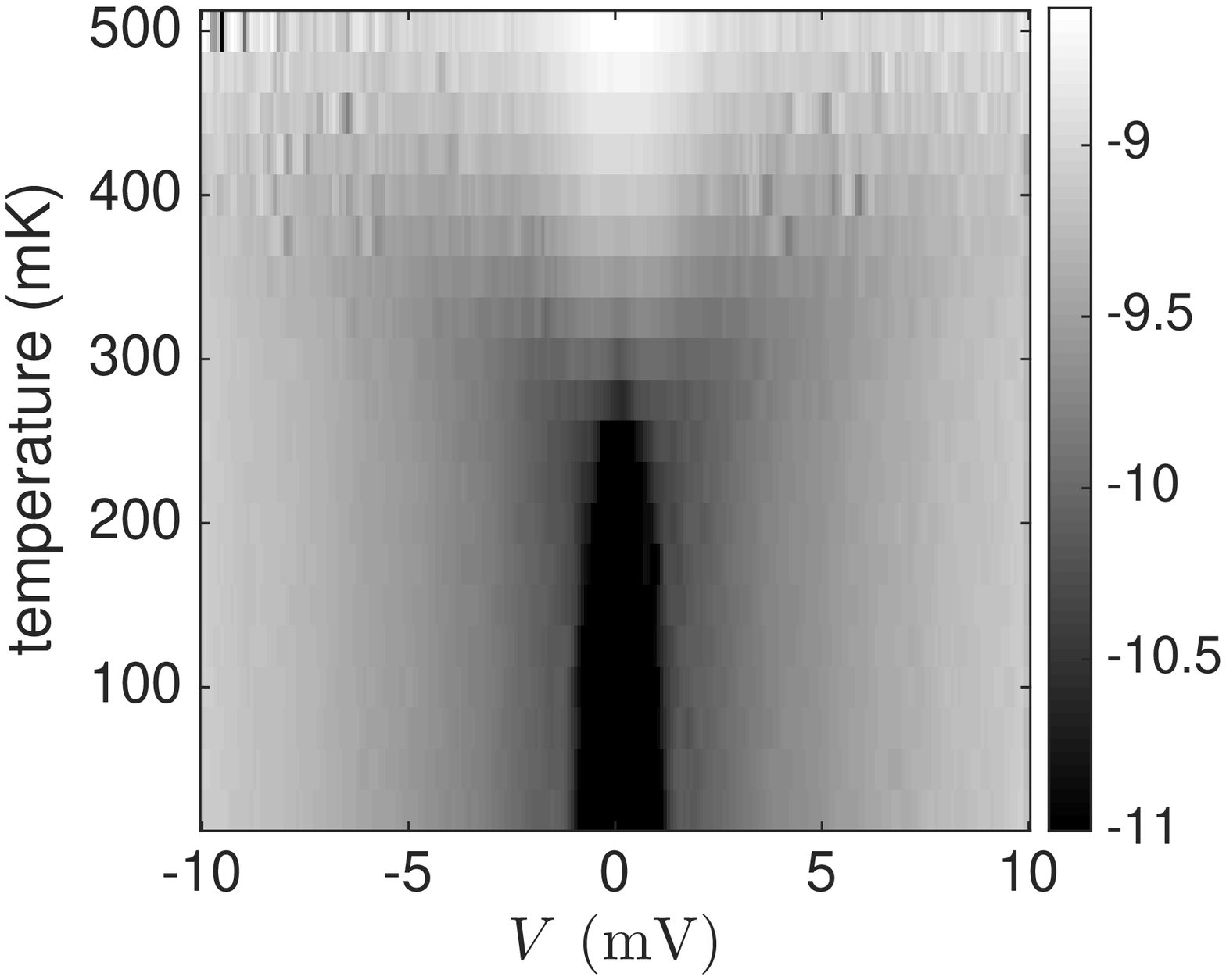}

   }\\
  \subfloat
   {
       \includegraphics[width=3.3in]{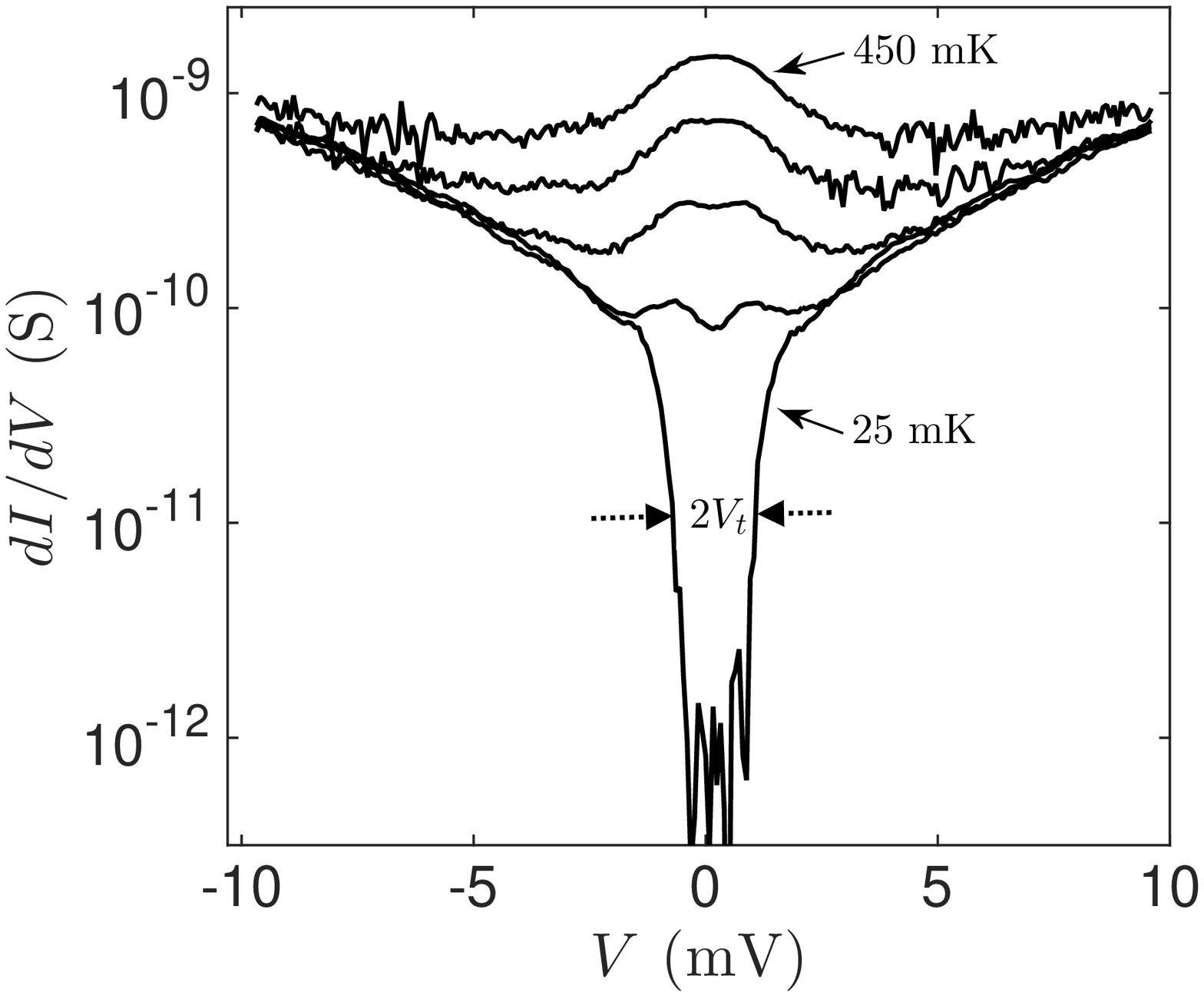}
         \label{fig:second_sub}
   }
        \caption{Differential conductance $dI/dV$ for device B in zero magnetic field. Upper: grayscale image of $dI/dV$ ($\Omega^{-1}$) on a logarithmic scale, showing a temperature dependent threshold that vanishes at a well-defined temperature, identified as the parity temperature $T^*$. Lower: slices of $dI/dV$ for temperatures 25, 300, 350, 400, and 450 mK. For temperatures below $T^*$, the threshold voltage is clearly distinguished as a nearly two orders of magnitude increase in the conductance, as shown by the 25mK data. The voltage threshold $V_t$ is experimentally determined to be the absolute value of the voltage bias at which the conductance rises above $10^{-11}\,\Omega^{-1}$, which is a factor of 10 greater than the measurement resolution  $G_0^\textrm{min}$\,=\,$10^{-12}\, \Omega^{-1}$.
}
    \label{fig:sample_subfigures}
\end{figure}

\begin{figure}
 \subfloat
      {
	 \includegraphics[width=3.3in]{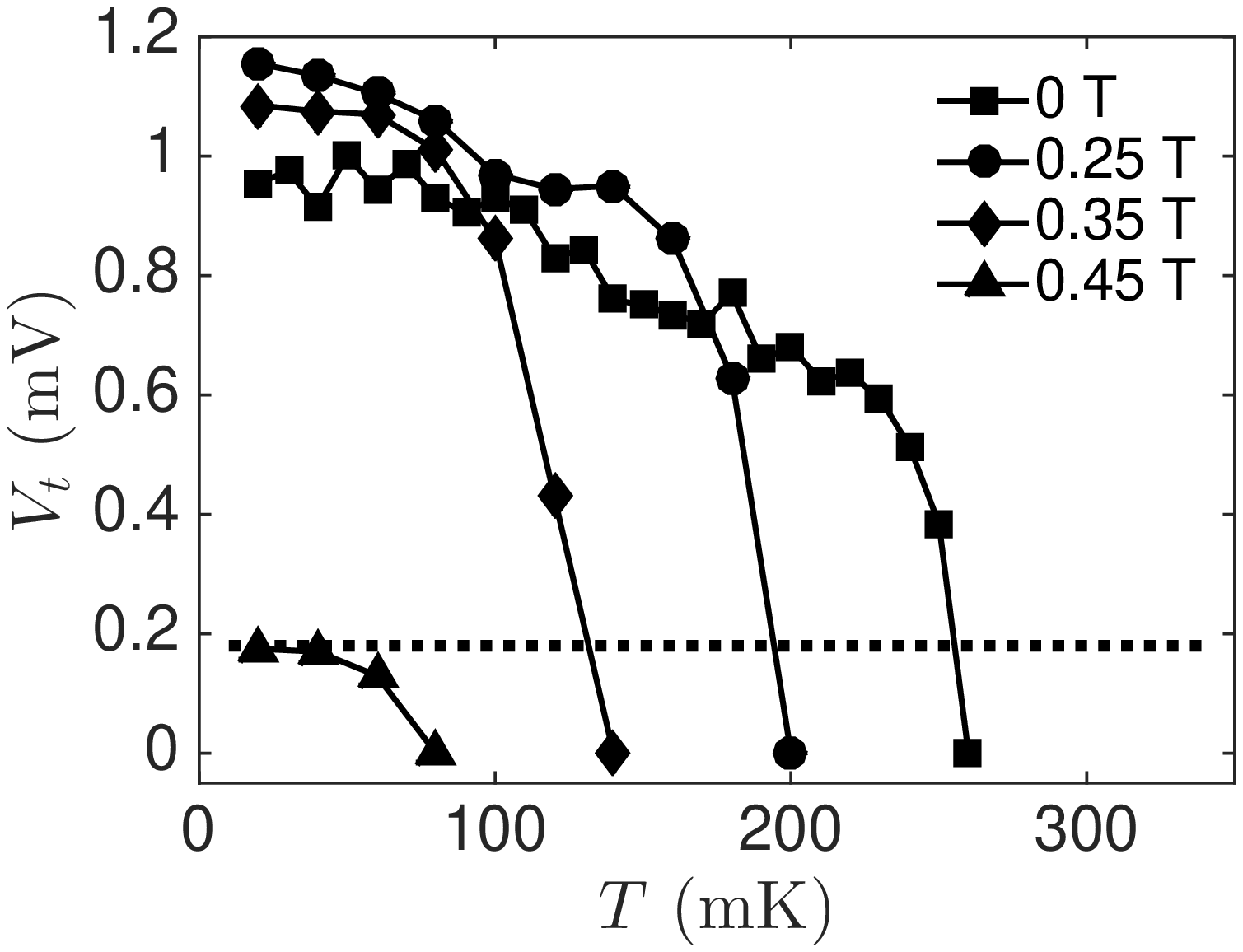}
      } \\
\subfloat
     {
        \includegraphics[width=3.3in]{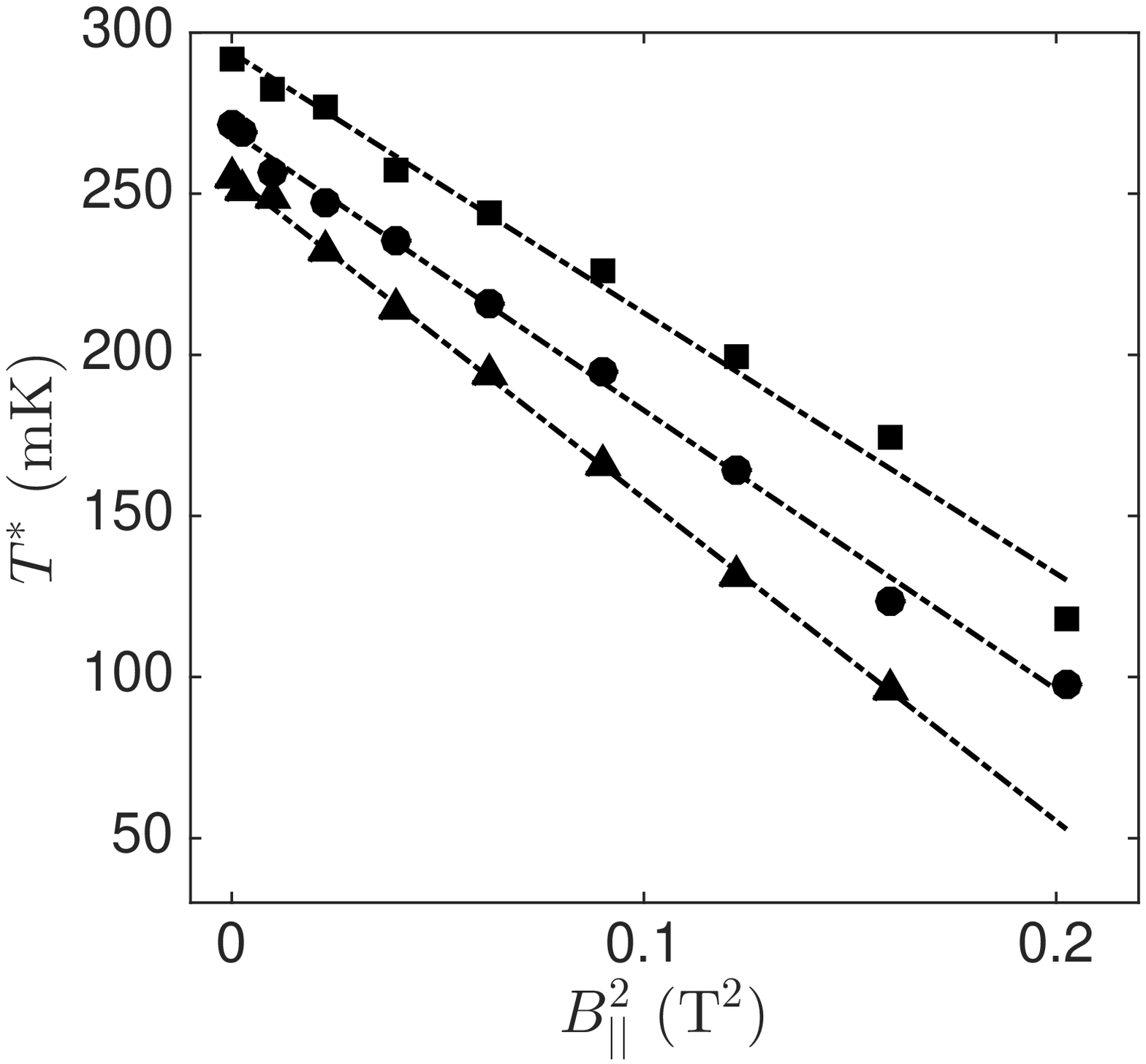}
}
  
        \caption{Upper plot: temperature dependence of the voltage threshold $V_t$ for device A for $B_{||}$ = 0, 0.25, 0.35 and 0.45 T. Lower plot:  magnetic field dependence of the parity temperature $T^*$ for devices A (triangles), B (circles) and C (squares). }
    \label{fig:tstar}
\end{figure}

\section{Parity effect}

The dependence of the measured threshold voltage $V_t$ on temperature for device A is shown in the upper plot of Figure 4 for several values of the parallel magnetic field, $B_{||}$= 0, 0.25, 0.35 and 0.45 T. It is evident that $V_t$ vanishes sharply at a specific temperature that depends on $B_{||}$. We argue that this behavior is a consequence of the parity effect for small superconducting islands. The ground state free energy for an odd number of electrons is higher than that for even numbers by the amount $F=\Delta-k_\mathrm{B}T \ln N_\mathrm{eff}$, where $N_\mathrm{eff} (T)\approx \mathcal{V}\rho(0)\sqrt{2\pi k_\mathrm{B}T\Delta(T)}$ is the effective number of states at finite $T$ arising from integration over the BCS quasiparticle DOS, $\mathcal{V}$ is the volume of the island and $\rho(0)$ is the density of states for the normal metal at the Fermi energy\cite{PhysRevLett.69.1993.Averin}. The free-energy difference $F$ for a single island vanishes at a cross-over temperature $k_\mathrm{B}T^* = \Delta / \ln N_\mathrm{eff}(T^*)$. Using the island volumes given in Table I, the experimentally determined $\Delta_0$, and taking  $\rho(0) = 1.45 \times 10^{47}$ m$^{-3}$ J$^{-1}$, for the density of states for aluminum,  one can compute the theoretically expected parity temperature for isolated islands. For islands such as those in devices A and B, we calculate $T^*$\,=\,260\,mK, and for the islands of device C,  $T^*$\,=\,277\,mK. One notices that for our device parameters, $k_\mathrm{B}T^* \approx \Delta /9$, which is much less than $T_C\,\approx$\,1.3\,K for aluminum.

We experimentally determine the array parity temperature $T^*$ as the temperature at which $V_t$ passes through the voltage threshold observed in the normal state at the lowest temperature, which is indicated in Figure 4 (upper plot) by the horizontal dotted line. For $B_{||}=0$ we find $T^*$\,=\,256, 269 and 294\,mK, respectively, for devices A, B and C.
The dependence of $T^*$ on $B_{||}^2$ is shown in Figure 4 for the three devices, and is observed to be linear with $B_{||}^2$, hence linear with $\Delta(B_{||})$ as expected.  Given that $\Delta(B_{||})$ is found to be nearly the same for all devices, the results imply a different $N_\mathrm{eff}$ for each device. This result can only partially be understood within the single-island picture of the parity effect. The island volume for device C is nearly twice that for devices A and B, (see Table 1). Island volume affects $T^*$ logarithmically, and therefore would only lead to a variation in $T^*$ of 17mK. However, the observed difference in $T^*$  between device B and C is 25mK, and between devices A and C, 38mK. Devices A and B, fabricated in the same evaporation and located in the same EBL field, show a difference in $T^*$ of 13\,mK, even though they have nominally the same island volume. 

The data suggest that the inferred $N_\mathrm{eff}$ for our devices does not follow from single-island considerations alone, but could be explained if the effective BCS quasiparticle DOS for the system of coupled islands is modified due to charge transport processes. One could expect  $N_\mathrm{eff}$ to depend on factors such as the charging energy, tunnel resistance, and offset charge and other disorder in the array. As can be seen in the large-scale $IVC$ data of Figure 2 (discussed above), device A with the lowest $T^*$ also exhibits the largest BCS peak in the tunneling DOS, followed by device B, which has the second lowest $T^*$. 
 
Although the parity effect in superconducting single-electron transistors has been well known for the last two decades, it has been almost entirely neglected in studies of Josephson-junction arrays. A sole theoretical paper by Feigel'man \textit{et al.}\cite{FeigelmanJETP} pointed out the implications of the parity effect on the experimental search for the charge-unbinding transition in 2D junction arrays. Discussions of the parity effect in experimental works on both 2D junction arrays and 2D disordered superconducting films, however, are conspicuously absent. Our results are particularly relevant for the latter, as the theoretical models describing disordered superconducting films below the superconductor-insulator transition are based on a junction-array picture for the mesoscopically-structured films\cite{Fistul_etal_2008}.

In voltage-biased single-electron transistors, where there is no threshold for conduction, the parity effect is observed by change in the periodicy of the gate-dependent current at finite bias\cite{PhysRevLett.69.1997.Tuominen,AmarPRL94}. In a Cooper-pair box, the parity effect can be seen directly by measuring the box charge and observing a transition from $1e$ to $2e$ periodic Coulomb staircases\cite{LafargeNature93, LafargePRL93}. In the array measurements presented here, however, the parity effect has a global effect on charge transport across the whole chain of junctions. For a single island, the parity-dependent free energy difference goes to zero precisely as the thermal expectation for the number of BCS quasiparticles $\left< N_{qp} \right>$ no longer depends on the charge parity of the island\cite{Tinkham_etal_95}. In our junction chains, it appears that the presence of $\sim$1 thermally-excited BCS quasiparticle per island effectively destroys the insulating state, by sharply removing the voltage threshold for single-electron injection. In addition, the tunneling rates for single electrons through the chain are significantly enhanced due to the presence of quasiparticles. The precise microscopic mechanism underlying this phenomena is currently being investigated. Some simulation results on parity effects in arrays can be found in Cole \textit{et al.}\cite{arXiv:1503.01905}. Recent experimental results on a hybrid normal-superconducting transistor illustrate enhanced charge tunneling due to a non-equilibrium quasiparticle distribution.\cite{HeimesPRB2014}

\section{Thermally-activated conductance}

\begin{figure}

 \subfloat
      {
	 \includegraphics[width=3.3in]{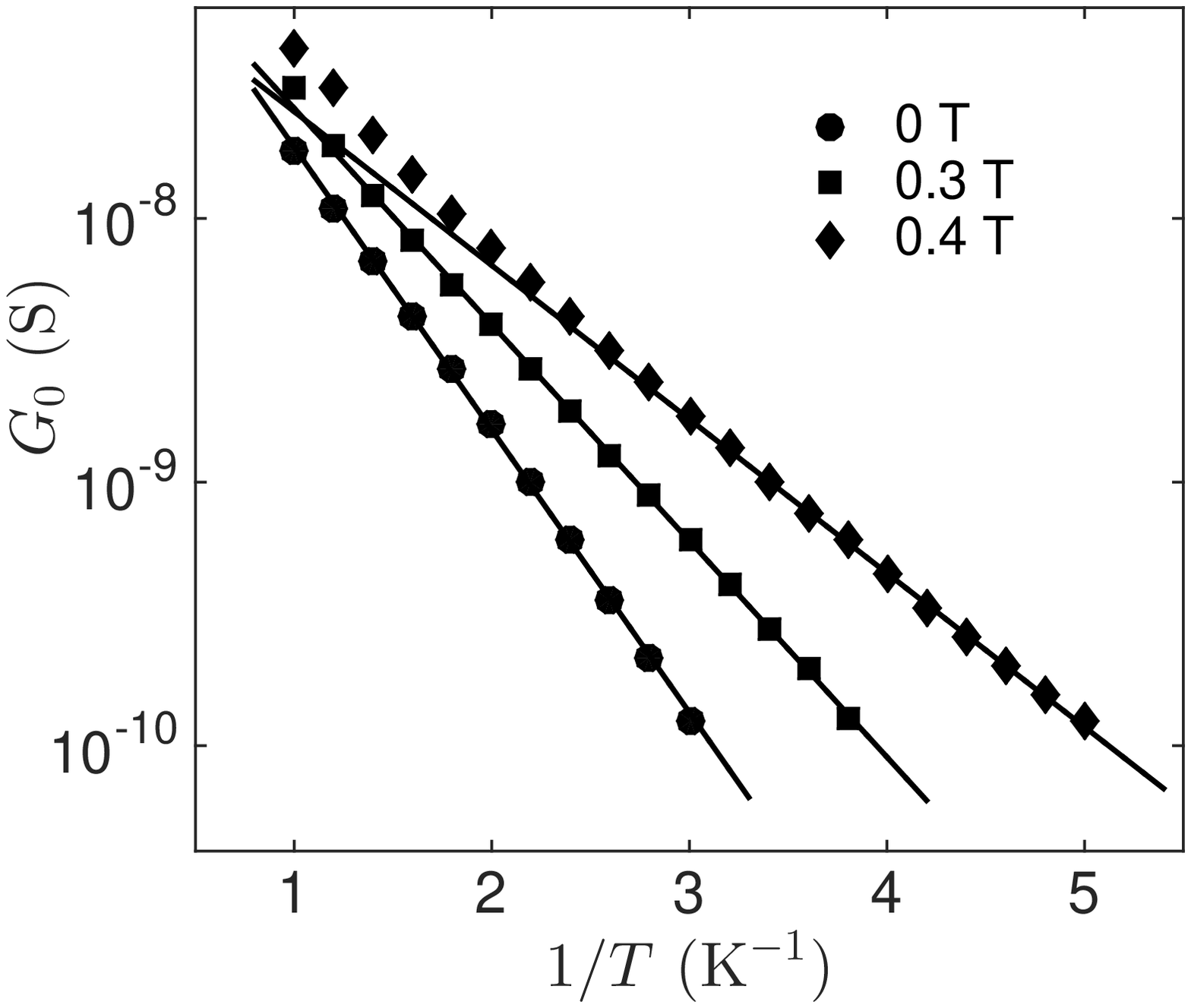}
      } \\    
   
 \subfloat    
    {
       \includegraphics[width=3.3in]{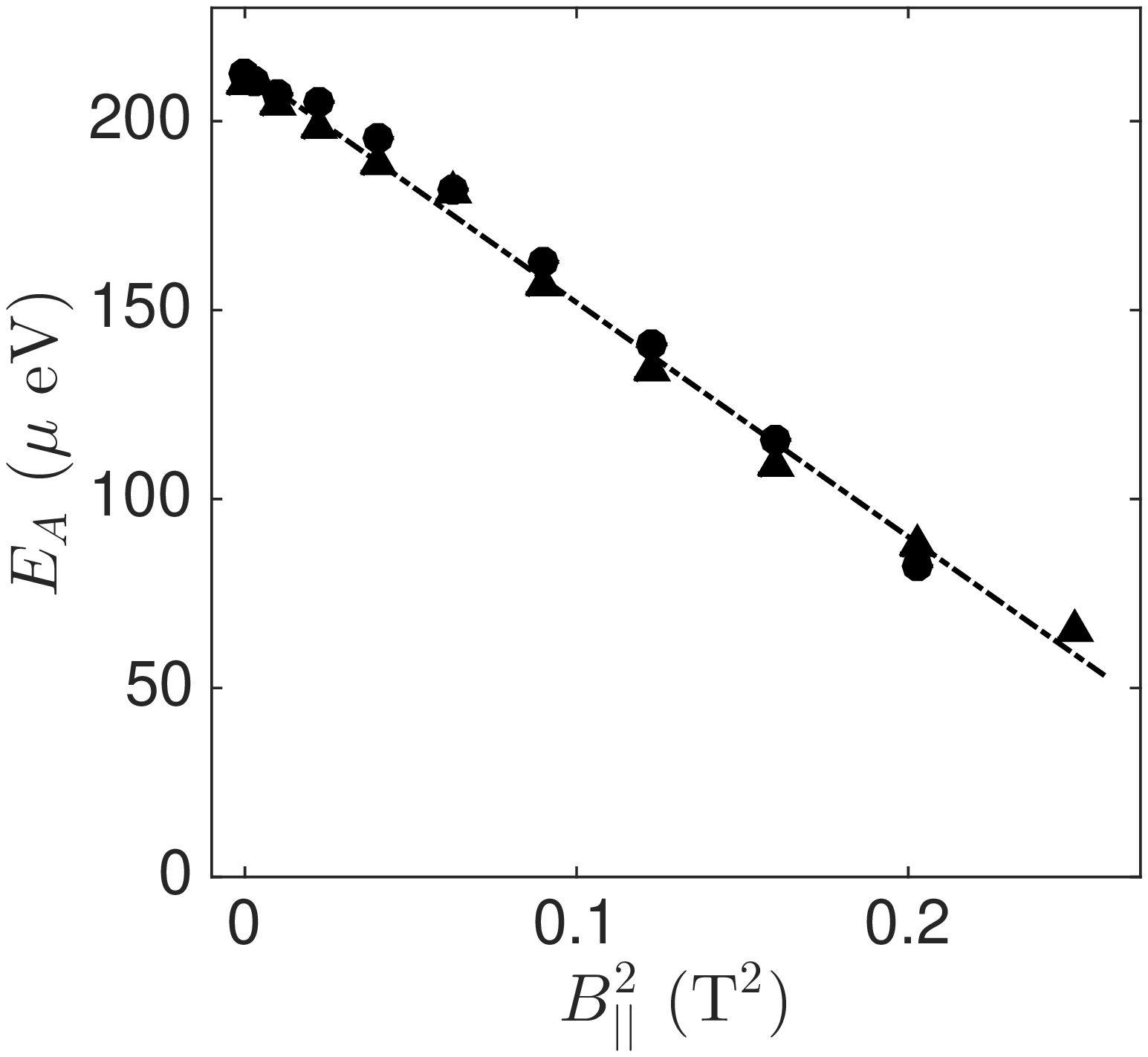}
         \label{fig:activation}
    }
        \caption{Upper plot:  log\,$G_0$ \textit{vs.} $1/T$ for device B for $B_{||}$=0, 0.3 and 0.4\,T. The solid lines are fits to an Arrhenius law. Lower plot: activation energy $E_A$ \textit{vs.}  $B_{||}^2$ for devices A (triangles) and B (circles). The dot-dashed line is a combined fit for both devices to $\Delta(B_{||})=\Delta_0(1-B^2_{||}/B^2_{c||})$ yielding zero-field gap $\Delta_0=214\pm 3\,\mu e \mathrm{V}$and $B_{c||}=0.59\pm 0.02$\,T.} 
    \label{fig:sample_subfigures}
\end{figure}

The zero-bias conductance $G_0\equiv (dI/dV)_{V=0}$ was measured under applied parallel magnetic fields for temperatures ranging from the parity temperature up to 1K. $G_0$ in all devices is found to follow an Arrhenius law for thermal activation, $G_0(T) = G_\infty\exp(-E_A/k_\mathrm{B}T)$, where $E_A$ is the activation energy, as shown in the upper plot of Figure 5. The zero-bias conductance in the normal state at 0.6\,T (not shown) continues to decrease above $T^{-1}$\,=\,10, indicating an electronic temperature lower than 100 mK\cite{Pekola94}.

As a function of applied $B$-field, the activation energy for devices A and B is linear with $B_{||}^2$, as shown in Figure 5, and appears to be equal to $\Delta(B_{||})$. The data can be accurately fit to $E_A=\Delta(B_{||})=\Delta_0(1-B^2_{||}/B^2_{c||})$, with $\Delta_0=214\pm3\mu eV$, and $B_{c||}=0.59\pm 0.02\,T$. We find that the experimentally determined value of $E_A$ in zero applied field agrees with an independent estimate of $\Delta_0$ gained from the large scale $IVC$ data, $eV_\mathrm{qp}/2N=210\pm 10 \mu eV$, where $V_\mathrm{qp}$ is the voltage that marks the onset of direct quasiparticle tunneling that occurring for  $eV_\mathrm{qp}/N\simeq2\Delta$. An activation energy that equals the superconducting gap can be easily understood because an energy of $2\Delta$ is required to  break a Cooper-pair. Since two independent excitations are created, the exponent for thermal activation is $\Delta$ rather than $2\Delta$.\cite{TinkhamBook}

For device C, $E_A$ varies randomly and somewhat irreproducibly with  $B_{||}$, taking values  between 250 and 350 $\mu e$V. We attribute this to significantly larger disorder present in device C, as inferred from the large scale IV data, and which is also consistent with the much larger $E_{CP}$ for device C. In contrast to this, however, the activation exponent for the conductance evaluated at $V=1.5$mV for device C shows nearly identical behavior to that of $G_0$ for devices A and B. This voltage bias is just outside the observed Coulomb blockade region of device C in the normal state, which is also relevant for unpaired charge carriers when the islands of the chain are superconducting. We conclude that transport above the parity temperature is set by thermally-activated quasiparticles in chains where strong charge disorder does not dominate.

Thermally-activated transport in 1D SQUID-arrays was reported recently by Zimmer  \textit{et al.}\cite{PhysRevB.88.144506.Zimmer}. The use of a SQUID geometry permitted tuning $E_J$ \textit{in situ} using a perpendicular magnetic field.  To account for their measurements, Zimmer \textit{et al.} assume a zero-bias conductance that is the sum of two contributions: a flux-dependent part, as would be expected for Cooper-pair tunneling, and a flux-independent term that remains when $E_J$ (Cooper-pair tunneling) becomes very small. With $E_J$ suppressed to nearly zero, Zimmer  \textit{et al.} observe an activation exponent of the order of the superconducting gap. These authors interpret their measured $E_A$ as a characteristic charging energy for localized Cooper pairs undergoing variable-range hopping, although they mention as an alternative explanation, thermally generated quasiparticles.

As far as we are aware, Zimmer  \textit{et al.} is the only reported measurement of thermally-activated zero-bias conduction in 1D arrays in the superconducting state. Thermally-activated zero-bias conductance in 2D junction arrays has been reported by two groups some time ago\cite{Tighe93,DelsingPRB94}. These authors interpreted their results in terms of a so-called ``core energy'', $E_\mathrm{core}$, which is  the energy required to create an electron-hole pair on adjoining sites (\textit{e.g.} by moving a single electron by one site), together with the induced polarization charge on neighboring islands. This model is known as the soliton model. In the superconducting state, one finds for a 2D system, $E_\mathrm{core}=2\Delta+E_C/2$, where the first term come from breaking a Cooper pair, and the second is the electrostatic energy for placing a single electron and hole on adjoining sites. Since two independent excitations are created, $E_A=E_\mathrm{core}/2=\Delta+E_C/4$. While Tighe \textit{et al.}\cite{Tighe93} found quantitative agreement with the core-energy model of localized dipoles, more detailed measurements by Delsing \textit{et al.}\cite{DelsingPRB94} showed substantial deviations from this picture. For large $\Delta/E_{CP}$, Delsing \textit{et al.}\cite{DelsingPRB94} interpreted their results as evidence for Cooper-pair/hole solitons, even though their measured activation energies showed a strong dependence on superconducting gap. Delsing \textit{et al.}\cite{DelsingPRB94} also report thermally-activated conduction in a 1D chain, but only in the normal state.

For a 1D chain with localized dipole excitations, one expects $E_\mathrm{core}=2\Delta+E_C$ using the soliton model of Tighe \textit{et al.} and Delsing \textit{et al.}\cite{Tighe93, DelsingPRB94} , and therefore $E_A=E_\mathrm{core}/2=\Delta+E_C/2=\Delta+E_{CP}/8$. Our results for 1D chains show that $E_A$ agrees more closely with $\Delta$, with no additional term needed to account for the charging energy of electron-hole pairs on adjacent islands. The localized dipole model ignores tunneling processes that effectively lower the core energy. As noted previously, above $T^*$, a voltage threshold for conduction is no longer found. In summary, conductance above $T^*$ is consistent with the lack of an electrostatic threshold for both charge injection and activated transport.

\begin{figure}
    \centering
     \includegraphics[width=3.3in]{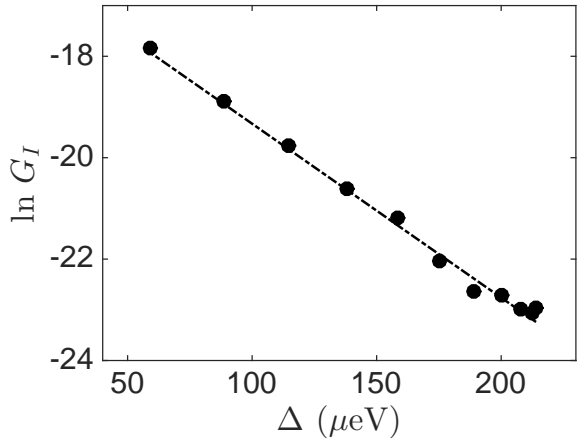}
      \label{fig:injection}
   \caption{Magnetic field dependence of the conductance in the injection regime for device A at $T$\,=\,20 mK. The conductance at  $V$\,=\,4\,mV, $G_I$, is plotted against the experimentally determined $\Delta(B_{||}^2)$. The dot-dashed line is a fit to the expression $\mathrm{ln}\,G_I=\mathrm{ln}\,G^0_I - \Delta/k_\mathrm{B}T_\mathrm{eff},$ which yields $T_\mathrm{eff}=340$\,mK.}
\end{figure}

\section{Conductance below the parity temperature}

 Finally, we have measured conductance at 20 mK and above the threshold voltage $V_t$ as a function of the magnetic field. Data for device A is shown in Figure 6 taken at a bias voltage, $V$\,=\,4\,mV. Here we have used the fit from Figure 5 to express $B_{||}$ in terms of $\Delta$. As the superconducting gap is suppressed by the magnetic field, the conduction $G_I$, in what we will call the ``injection regime'', is clearly exponentially enhanced by the factor $\exp\left( -\Delta / k_\mathrm{B} T_\mathrm{eff} \right)$.   In contrast to the zero-bias conductance, we find the effective temperature  $T_\mathrm{eff}=340$\,mK, which is considerably larger than the zero-field parity temperature for this device, $T^*=260$\,mK. This shows that charge transport below $T^*$ and above the voltage threshold occurs by injection of single electrons/holes into a non-equilibrium steady state, which shows a significantly elevated effective temperature. Future experiments are needed to address the detailed nature of this steady state and it's relation to the voltage threshold for conduction, $V_t$, observed at low temperatures.

\section{Conclusion}
 
In conclusion, we find that for 1D Josephson-junction chains deep in the insulating regime, where $E_J \ll E_{CP}$, there is a characteristic parity temperature $T^*$, above which the insulating state is destroyed by thermally-excited BCS quasiparticles.  Above $T^*$, an observable zero-bias conductance appears and is thermally-activated with an activation energy equal to the superconducting gap. This can be understood most simply if charge carriers are single electrons and holes rather than Cooper pairs. The situation is somewhat analogous to donor ionization in a doped semiconductor, although here the donors are the background of localized Cooper pairs, and the effective ionization energy is strongly renormalized due to the singular BCS density of states. Conduction at temperatures below $T^*$ occurring above the threshold voltage appears to be thermally-activated, with an exponent equal to the ratio of the superconducting gap to an effective thermal energy, $k_BT_\mathrm{eff}$. The effective temperature $T_\mathrm{eff}$ is found to be significantly higher than the electronic temperature that would otherwise exist in the array. This indicates that a non-equilibrium steady state of unpaired charge carriers becomes established, enabling above-threshold charge transport below the parity temperature in the Cooper-pair insulator. Our results are also relevant to studies of disordered superconducting films, which are often modeled using a picture of weakly coupled superconducting islands.

\section{Acknowledgments}
 
This work was supported by the Centre of Excellence for Engineered Quantum
Systems, an Australian Research Council Centre of Excellence, CE110001013. JHC is supported by the Victorian Partnership for Advanced Computing (VPAC). We would like to acknowledge helpful discussions with Michael Marthaler at Karlsruhe Institute of Technology. Devices were fabricated at the UNSW Node of the Australian National Fabrication Facility (ANFF), and we would like to acknowledge the expert advice and help of ANFF staff in device processing.


\begin{thebibliography}{99}

\bibitem{PhysRevB.30.1138.Bradley} 
	R.~M.~Bradley, S.~Doniach,
	\newblock Phys. Rev. B \textbf{30}, 1138
	(1984).

\bibitem{PhysRep.355.235.Fazio} 
	Rosario\,Fazio and Herre\,van\,der\,Zant,
	\newblock Physics Reports \textbf{355}, 235
	(2001), and references therein.


\bibitem{PhysRevLett.79.3736.Glazman}
	L.~I.~Glazman and A.~I.~Larkin,
	\htmladdnormallink{\newblock Phys. Rev. Lett. \textbf{79}, 127001 (1997)} {http://journals.aps.org/prl/abstract/10.1103/PhysRevLett.79.3736}.

\bibitem{PhysRevLett.88.227005.Doucot}
	Benoit Dou{\c c}ot and Julien Vidal,
	\htmladdnormallink{\newblock Phys. Rev. Lett. \textbf{88}, 227005 (2002)} {http://journals.aps.org/prl/abstract/10.1103/PhysRevLett.88.227005}.


\bibitem{PhysRevLett.90.107003.Doucot}
	B.~Dou{\c c}ot, M.~V.~Feigel'man, and L.~B.~Ioffe,
	\htmladdnormallink{\newblock Phys. Rev. Lett. \textbf{90}, 107003 (2003)} {http://journals.aps.org/prl/abstract/10.1103/PhysRevLett.90.107003}.


\bibitem{Nature.415.503.Ioffe} 
	L. B. Ioffe, M. V. Feigel'man, A. Ioselevic, D. Ivanov, M. Troyer, and G. Blatter,
	\htmladdnormallink{\newblock Nature \textbf{415}, 503 (2002).
	}{http://www.nature.com/nature/journal/v415/n6871/abs/415503a.html}
	


\bibitem{IntJModPhysB.24.4081.Goldman} 
	A.~M.~Goldman,
	\htmladdnormallink{ \newblock Int. J. Mod. Phys. B \textbf{24}, 4081
	(2010)}{http://www.worldscientific.com/doi/abs/10.1142/S0217979210056451}, and references therein.


\bibitem{NaturePhys.6.589.Pop} 
	I. M. Pop, I. Protopopov, F. Lecocq, Z. Peng, B. Pannetier, O. Buisson, and W. Guichard,
	\htmladdnormallink{\newblock Nature Physics \textbf{6}, 589 (2010).
	}{http://www.nature.com/nphys/journal/v6/n8/abs/nphys1697.html}


\bibitem{PhysRevB.85.024521.Manucharyan}
	Vladimir E. Manucharyan, Nicholas A. Masluk, Archana Kamal, Jens Koch, Leonid I. Glazman, and Michel H. Devoret,
	\htmladdnormallink{\newblock Phys. Rev. B \textbf{85}, 024521 (2012)} {http://journals.aps.org/prb/abstract/10.1103/PhysRevB.85.024521}.


\bibitem{NewJPhys.15.095014.Ergul}
	Adem Erg\"ul, Jack Lidmar, Jan Johansson, Yağız Azizo{\v g}lu, David Schaeffer, and David B. Haviland,
	\htmladdnormallink{\newblock New Journal of Physics \textbf{15}, 095014 (2013)} {http://iopscience.iop.org/1367-2630/15/9/095014}.



\bibitem{Nature.434.361.Bylander} 
	Jonas Bylander, Tim Duty, and Per Delsing, 
	\newblock Nature \textbf{434},
	\htmladdnormallink{361}{http://www.nature.com/nature/journal/v434/n7031/full/nature03375.html}
	(2005).


\bibitem{PhysRevB.54.6857R.Haviland}
	David B.~Haviland and Per Delsing,
	\htmladdnormallink{\newblock Phys. Rev. B \textbf{54}, 6857(R) (1996)} {http://journals.aps.org/prb/abstract/10.1103/PhysRevB.54.R6857}.


\bibitem{JLTP.124.291.Agren}
	Peter \AA gren, Karin Andersson and David B. Haviland
	\htmladdnormallink{\newblock J. Low Temp. Phys. \textbf{124}, 291 (2001)} {http://link.springer.com/article/}.


\bibitem{PhysRevB.83.064517}
	Jens Homfeld, Ivan Protopopov, Stephan Rachel, and Alexander Shnirman,
	\htmladdnormallink{\newblock Phys. Rev. B \textbf{83}, 064517 (2011).} {http://journals.aps.org/prb/abstract/10.1103/PhysRevB.83.064517}

\bibitem{Masluk_etal_2012} 
Nicholas~A.~Masluk, Ioan~M.~Pop, Archana~Kamal, Zlatko~K.~Minev, and Michel~H.~Devoret,
\htmladdnormallink{\newblock Phys. Rev. Lett. \textbf{109}, 137002 (2012).} 
{http://link.aps.org/doi/10.1103/PhysRevLett.109.137002}



\bibitem{PhysRevB.76.020506R.Bylander} 
	Jonas Bylander, Tim Duty, G\"oran Johansson, and P. Delsing, 
	\newblock Phys. Rev. B  \textbf{76},  020506(R)
	(2007).

\bibitem{NewJPhys.16.063019.Cole} 
	Jared H.~Cole, Juha Lepp\"akangas, and Michael Marthaler,
	New J. Phys.  \textbf{16}, 
	\htmladdnormallink{063019}{http://iopscience.iop.org/1367-2630/16/6/063019/}
	(2014).


\bibitem{PhysRevLett.103.127001.Syzranov}
	S.V. Syzranov, K.B. Efetov and B.L. Altshuler,
	\newblock Phys. Rev. Lett. \textbf{103},
	\htmladdnormallink{127001}{http://link.aps.org/doi/10.1103/PhysRevLett.103.127001}
	(2009).


\bibitem{Fistul_etal_2008}
M.~V.~Fistul, V.~M.~Vinokur, and T.~I.~Baturina,
\newblock Phys. Rev. Lett. \textbf{100},
\htmladdnormallink{086805}{http://link.aps.org/doi/10.1103/PhysRevLett.100.086805}



\bibitem{PhysRevB.88.144506.Zimmer}
	J. Zimmer, N. Vogt, A. Fiebig, S.V. Syzranov, A. Lukashenko, R. Sch\"afer, H. Rotzinger, A. Shnirman, M. Marthaler and A.V. Ustinov,
	\newblock Phys. Rev. B \textbf{88},
  	\htmladdnormallink{144506}{http://link.aps.org/doi/10.1103/PhysRevB.88.144506}, (2013.)

\bibitem{ChowPRL98}
	Edmond~Chow, Per~Delsing and David~B.~Haviland,
	\newblock Phys. Rev. Lett \textbf{81}, 
	\htmladdnormallink{204}{http://link.aps.org/doi/10.1103/PhysRevLett.81.204}, (1998).
	
	
\bibitem{CWang2014}
	C.~Wang, Y.~Y.~Gao, I.~M.~Pop, U.~Vool, C.~Axline, T.~Brecht, R.~W.~Heeres, L.~Frunzio, M.~H.~Devoret, G.~Catelani, L.~I.~Glazman, and R.~J.~Schoelkopf,
	\newblock Nat. Commun. \htmladdnormallink{5:5836}{doi:10.1038/ncomms6836} (2014).

\bibitem{Aumentado2004}
	J.~Aumentado, M.~W.~Keller, J.~M.~Martinis, and M.~H.~Devoret,
	\newblock Phys. Rev. Lett. \textbf{92},
	\htmladdnormallink{066802}{http://link.aps.org/doi/10.1103/PhysRevLett.92.066801}, (2004).
	
\bibitem{arXiv:1503.01905}
	Jared\,H.\,Cole, Andreas\,Heimes, Timothy\,Duty, and Michael\,Marthaler, 
	\newblock Phys. Rev. B \textbf{91}, \htmladdnormallink{184505}{http://link.aps.org/doi/10.1103/PhysRevB.91.184505}, (2015).
	
\bibitem{PhysRevLett.69.1993.Averin}
	D.V. Averin and Yu.V. Nazarov,
	\newblock Phys. Rev. Lett. \textbf{69},
	\htmladdnormallink{1993}{http://link.aps.org/doi/10.1103/PhysRevLett.69.1993}, (1992).

\bibitem{PhysRevLett.69.1997.Tuominen}
	M.T. Tuominen, J.M. Hergenrother, T.S. Tighe and M. Tinkham,
	\newblock Phys. Rev. Lett. \textbf{69},
	\htmladdnormallink{1997}{http://link.aps.org/doi/10.1103/PhysRevLett.69.1997}, (1992).

\bibitem{AmarPRL94}
	A.~Amar, D.~Song, C.~J.~Lobb, and F.~C.~Wellstood,
	\newblock Phys. Rev. Lett. \textbf{72},
	\htmladdnormallink{3234}{http://link.aps.org/doi/10.1103/PhysRevLett.72.3234}, (1994).
	

\bibitem{LafargePRL93}
	P.\,Lafarge, P.\,Joyez, D.\,Esteve, C.\,Urbina, and M.\,H.\,Devoret.
	\newblock Phys. Rev. Lett. \textbf{70},
	\htmladdnormallink{994}{http://link.aps.org/doi/10.1103/PhysRevLett.70.994}, (1993).

\bibitem{LafargeNature93} P.~Lafarge,~P. Joyez, C.~Urbina, and M.~H.~Devoret,
	\newblock Nature \textbf{365}, 422 (1993).

\bibitem{TinkhamBook}\textit{Introduction to superconductivity}, Michael Tinkham, 
Dover, New York (2004), second edition.

\bibitem{FeigelmanJETP}
M.~V.~Feigel'man, S.~E. Korshunov, and A. B. Pugachev,
 \newblock Soviet Physics JETP \textbf{65}, 566 (1997).

\bibitem{Tinkham_etal_95}
	M.~Tinkham, J.~M.~Hergenrother, and J.~G.~Lu,
       \newblock Phys. Rev. B \textbf{51},
	\htmladdnormallink{12649}{http://link.aps.org/doi/10.1103/PhysRevB.51.12649}, (1995).



\bibitem{HeimesPRB2014}
	Andreas~Heimes, Ville~F.~Maisi, Dmitri~S.~Golubev, Michael~Marthaler, Gerd~Sch\"on, and~Jukka~P.~Pekola,     		\newblock Phys. Rev. B \textbf{89},
	\htmladdnormallink{014508}{http://link.aps.org/doi/10.1103/PhysRevB.89.014508}, (2014).


	
\bibitem{Pekola94}
	J.P. Pekola, K.P. Hirvi, J.P. Kauppinen, and M. A. Paalanen, \newblock Phys. Rev. Lett. \textbf{73},
	\htmladdnormallink{2903}{http://link.aps.org/doi/10.1103/PhysRevLett.73.2903}, (1994).
	
\bibitem{Tighe93}
	T.~S.~Tighe, M.~T.~Tuominen, J.~M.~Hergenrother, and M.~Tinkham,
	\newblock Phys. Rev. B \textbf{47}, 
	\htmladdnormallink{1145}{http://link.aps.org/doi/10.1103/PhysRevB.47.1145}, (1993).

\bibitem{DelsingPRB94}
	P.\,Delsing, C.\,D.\,Chen, D.\,B.\,Haviland, Y.~\,Harada, and T.\,Claeson,
	\newblock Phys. Rev. B \textbf{50}, 
	\htmladdnormallink{3959}{http://link.aps.org/doi/10.1103/PhysRevB.47.3959}, (1994).


\end{thebibliography}

\end{document}